\documentclass[prl,aps,superscriptaddress,twocolumn,showpacs]{revtex4-1}
\usepackage{amsfonts}
\usepackage{amsmath}
\usepackage{amssymb}
\usepackage{graphicx}
\usepackage{color}
\usepackage{soul}	

\begin{document}

\title{Noncritical generation of nonclassical frequency combs\\
via spontaneous rotational symmetry breaking}
\author{Carlos Navarrete-Benlloch}
\affiliation{Max-Planck-Institut f\"{u}r die Physik des Lichts, Staudtstrasse 2, 91058 Erlangen, Germany}
\affiliation{Institute for Theoretical Physics, Erlangen-N\"urnberg Universit\"at, Staudtstrasse 7, 91058 Erlangen, Germany}
\affiliation{Max-Planck-Institut f\"{u}r Quantenoptik, Hans-Kopfermann-strasse 1, 85748 Garching, Germany}

\author{Giuseppe Patera}
\affiliation{Univ. Lille, CNRS, UMR 8523 - PhLAM - Physique des Lasers Atomes et Mol\'ecules, F-59000 Lille, France}

\author{Germ\'{a}n J. de Valc\'{a}rcel}
\affiliation{Departament d'\`{O}ptica, Facultat de F\'{\i}sica, Universitat de Val\`{e}ncia, Dr. Moliner 50, 46100 Burjassot, Spain}

\begin{abstract}
Synchronously pumped optical parametric oscillators (SPOPOs) are optical cavities containing a nonlinear crystal capable of down-converting a frequency comb to lower frequencies. These have received a lot of attention lately, because their intrinsic multimode nature makes them compact sources of quantum correlated light with promising applications in modern quantum information technologies. In this work we show that SPOPOs are also capable of accessing the challenging but interesting regime where spontaneous symmetry breaking plays a crucial role in the quantum properties of the emitted light, difficult to access with any other nonlinear optical cavity. Apart from opening the possibility of studying experimentally this elusive regime of dissipative phase transitions, our predictions will have a practical impact, since we show that spontaneous symmetry breaking provides a specific spatiotemporal mode with perfect squeezing for any value of the system parameters, turning SPOPOs into robust sources of highly nonclassical light above threshold.
\end{abstract}

\maketitle

\textbf{Introduction.--} To date, optical parametric oscillators (OPOs) constitute the main source of nonclassical states of light, finding numerous applications in emerging quantum technologies, e.g. in the fields of quantum metrology \cite{Taylor16,Treps02,Treps03,Taylor13,Taylor14,McKenzie02,Vahlbruch05,Abadie11,Grote13} or quantum information with continuous variables \cite{Braunstein05,Weedbrook12,NavarreteBook}. OPOs are optical cavities containing nonlinear crystals supporting the so-called parametric down conversion (PDC) process, by means of which a pump photon of frequency $\omega_\mathrm{p}$ is converted into a pair of photons of frequencies $\omega_\mathrm{s}$ and $\omega_\mathrm{i}$ (so-called, arbitrarily in the OPO case, signal and idler), and vice-versa, with $\omega_\mathrm{s}+\omega_\mathrm{i}=\omega_\mathrm{p}$ \cite{BoydBook,NavarretePhDthesis}. This generates strong quantum correlations between signal and idler (e.g. twin beams \cite{Reynaud87,Heidmann87}) The PDC Hamiltonian reads $\chi_\mathrm{p,s,i}f_\mathrm{p,s,i}\hat{a}_\mathrm{p}\hat{a}_\mathrm{s}^\dag\hat{a}_\mathrm{i}^\dag+\mathrm{H.c.}$ \cite{NavarretePhDthesis}, where $\hat{a}_j$ annihilate photons of the corresponding cavity modes, $\chi_\mathrm{p,s,i}$ is a coupling constant proportional to the nonlinear susceptibility and to the spatial overlap between the modes inside the nonlinear crystal, and $f_\mathrm{p,s,i}=\text{sinc}\hspace{0.5mm}\phi\equiv\frac{\sin\phi}{\phi}$, with $\phi=\frac{1}{2}(k_\mathrm{p}-k_\mathrm{s}-k_\mathrm{i})h$ the phase mismatch, being $h$ the crystal length and $k_j=n(\omega_j)\omega_j/c$ the wavenumber inside the crystal whose refractive index at frequency $\omega$ is $n(\omega)$. The condition $f=1$ ($\phi=0$, perfect phase matching) maximizes PDC, and usually selects which pair of signal-idler modes are efficiently generated.

Traditionally OPOs are operated under monochromatic pumping (a single $\omega_\mathrm{p}$). Since the parametric gain must compensate for cavity loss, a main feature of OPOs in the classical limit is the existence of a pumping threshold below which there is no emission, while above it a macroscopic field is excited in one specific signal-idler couple. In contrast, a fully quantum-mechanical theory accounts for the generation of photon pairs for any signal-idler couple satisfying energy and momentum conservation, even below threshold. However signal-idler modes of different couples do not show quantum correlations among them because the pump provides no appreciable feedback, since below threshold it is almost undepleted, while above the threshold its intracavity amplitude gets clamped to its threshold value \cite{Navarrete09,NavarretePhDthesis}.


\begin{figure}[t]
\begin{center}
\includegraphics[width=\columnwidth]{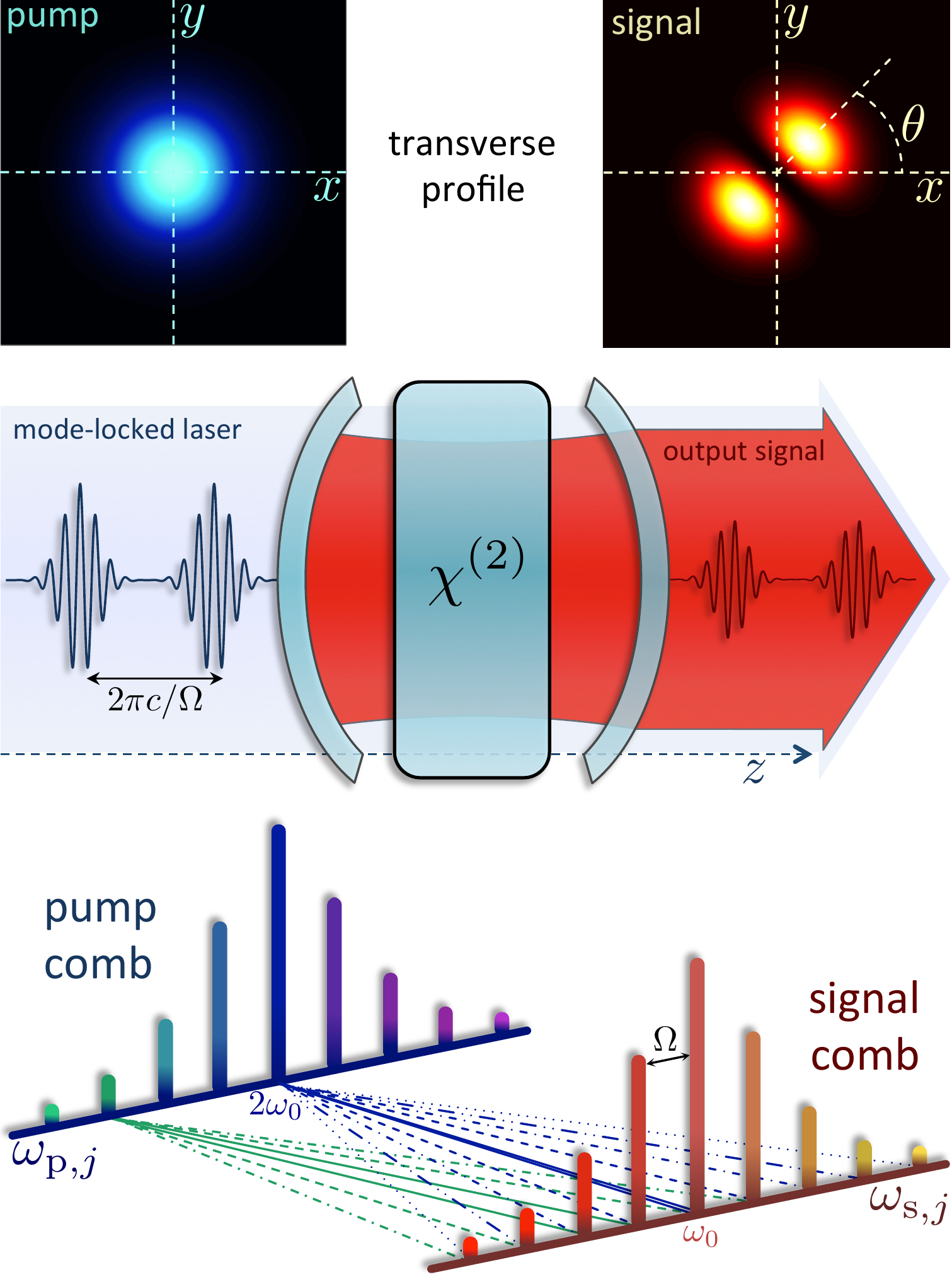}
\end{center}
\caption{(Center) OPO pumped by a mode-locked laser with repetition time $2\pi/\Omega$ equal to the OPO roundtrip time. The cavity is transparent for the pump and tuned to the first transverse mode family at the subharmonic. (Top) The pump beam has a Gaussian transverse profile, but classical down-conversion takes place in a TEM$_{10}$ mode with arbitrary orientation $\theta$. (Bottom) Pump and signal frequency combs with spectral-line spacing $\Omega$, and some of the down-conversion channels of two of the pump lines, $j=0$ and $j=-3$.}
\label{fig-Sketch}
\end{figure}

The situation changes dramatically when the pump comes from a mode-locked laser whose cavity roundtrip time matches the OPO one, $t_{\mathrm{cav}}$: Synchronously Pumped OPOs, or SPOPOs, see Fig. \ref{fig-Sketch}. Such multimode pump field consists of an infinite train of identical coherent pulses separated by $t_{\mathrm{cav}}$, known alternatively as a \textit{frequency comb} as its spectrum consists of discrete spectral lines separated by $\Omega=2\pi/t_\mathrm{cav}$, so-called cavity free spectral range. As in the monochromatic pump case, each pump spectral line generates multiple couples of signal-idler photons and, what is the new key ingredient, any signal/idler photon can be created by any of the different pump lines, what leads to massive quantum correlations between signal-idler photons at different frequencies. Special interest are receiving degenerate type I SPOPOs \cite{Valcarcel06,Patera10}, in which signal and idler have the same linear polarization and the perfect phase matching condition ($\phi=0$) happens for $\omega_\mathrm{s}=\omega_\mathrm{i}=\omega_\mathrm{p}/2\equiv\omega_0$, a condition achieved by proper crystal orientation and/or temperature tuning. In this case a single frequency comb around the subharmonic frequency $\omega_{0}$ is generated and the distinction between signal and idler photons is completely superfluous; hence we will refer to any subharmonic photon as a signal photon for brevity. In order to understand why such SPOPOs have so remarkable quantum properties, let us analyze their interaction Hamiltonian which can be written as \cite{Valcarcel06,Patera10}
\begin{equation} \label{Hint0}
\hat{H}=\mathrm{i}\hbar \chi\sum_{m,q}f_{m,q}\hat{p}_{m+q}\hat{s}_m^\dag\hat{s}_q^\dag+\mathrm{H.c.},
\end{equation}
where in the following we denote by $\hat{p}_j$ the annihilation operator of a pump photon of frequency $\omega_{\mathrm{p},j}=2\omega_0+j\Omega$, and $\hat{s}_j$ the annihilation operator of a signal photon of frequency $\omega_{\mathrm{s},j}=\omega_0+j\Omega$, with $j\in\mathbb{Z}$. The quantity $f_{m,q}\equiv\text{sinc}\left[\frac{1}{2}(k_{\mathrm{p},m+q}-k_{\mathrm{s},m}-k_{\mathrm{s},q}) h\right]$ is the phase-mismatch factor, and we assumed a common value for the coupling constant $\chi$ for any PDC channel, which is an excellent approximation \cite{Valcarcel06,Patera10}. Note that (\ref{Hint0}) is just the sum of infinite PDC Hamiltonians, each corresponding to the generic PDC channel $[2\omega_0+(m+q)\Omega] \rightarrow [\omega_0+m\Omega]+[\omega_0+q\Omega]$, and since this condition is verified for any $m,q\in\mathbb{Z}$, now all the signal modes are correlated with each other. This has enormous consequences: on one hand classical correlations appear (all signal modes get phase-locked, giving rise to well-defined trains of identical pulses separated by $t_\mathrm{cav}$), and on the other hand strong quantum correlations are built up, leading to highly multimode squeezing \cite{Valcarcel06,Patera10} and multipartite entanglement \cite{Patera12,Ferrini13}. Indeed, recent experiments have proven SPOPOs to be highly versatile sources of nonclassical light \cite{Medeiros14,Gerke15,Gerke16,Cai16}, with foreseen applications in quantum computation \cite{Ferrini13,Cai16} and communication \cite{Cai16}.

A neat way of analyzing the quantum dynamics of SPOPOs below threshold is by introducing the so-called ``supermodes'' \cite{Valcarcel06,Patera10,Patera12}, coherent superposition of cavity modes, which are special frequency combs that diagonalize the nonlinear interaction in which clean quantum properties are concentrated. Defining the annihilation operators $\hat{S}_k$ for these supermodes, Hamiltonian (\ref{Hint0}) becomes $\hat{H}=\mathrm{i}\hbar\chi\sum_k\Lambda_k\hat{S}_k^{\dag 2}+\mathrm{H.c.}$ (see the paragraph before Eq. (\ref{ClassE}) or \cite{Valcarcel06,Patera10,Patera12} for a definition of $\Lambda_k$), meaning that a degenerate SPOPO below threshold is just a collection of independent degenerate squeezers, but the modes that get squeezed are not individual cavity modes, but rather the supermodes. Note that $\hat{H}$ has the discrete symmetry $\hat{S}_k \rightarrow -\hat{S}_k$, meaning that the emission in a supermode is phase-locked but this locking is bistable, between two opposite phase values, exactly as degenerate OPOs \cite{WallsMilburnBook}, which is a signature of degenerate operation.

The strong multimode quantum field generated by SPOPOs is \textit{critical} in the sense that its nonclassicality is maximized at threshold, but it is rapidly degraded as the system is pumped further \cite{WallsMilburnBook,NavarretePhDthesis}, exactly as it happens with any nonlinear optical cavity where squeezing is linked with the presence of bifurcations \cite{Collet85}. Extending those features above threshold will improve the performance and reliability of these sources.

The Hamiltonian (\ref{Hint0}) describes the usual case of a degenerate SPOPO in which both the pump and signal modes have a Gaussian transverse profile. Recently, however, it has been predicted in the context of a degenerate OPO pumped by a monochromatic Gaussian beam, that when the cavity is tuned to the first transverse mode family at the subharmonic frequency \cite{Lassen09,Janousek09,Chalopin10}, the signal field displays a level of squeezing above threshold which equals that at threshold \cite{Navarrete08,Navarrete10,NavarretePhDthesis}, i.e., the squeezing production in this case is \textit{noncritical} \cite{Perez06,Perez07,Navarrete08,Navarrete09,Garcia09,Navarrete10,Garcia10}. The physics behind such a remarkable result lies in the spontaneous rotational symmetry breaking around the cavity axis brought about by the (above-threshold) classical field, which necessarily has the shape of a TEM$_{10}$ mode of arbitrary orientation because of orbital angular momentum (OAM) conservation (see Fig. \ref{fig-Sketch}). It follows that the TEM$_{01}$ mode orthogonal to this field has perfect quadrature squeezing at any operating point above threshold \cite{Navarrete08,Navarrete10,NavarretePhDthesis}. The main problem with OPOs under monochromatic pump is that, in practice, it is not possible to ensure that phase-matching is maximized for the degenerate process \cite{Harris69,Eckardt91,Fabre97}, what results in the oscillation a pair of non-degenerate modes above threshold. However this is not the case in SPOPOs even when the degenerate process is not perfectly phase-matched: As recent experiments have shown in the context of using SPOPOs as coherent Ising machines \cite{Marandi14,Takata16,McMahon16,Inagaki16}, they are truly degenerate above threshold, the emission displaying bistable phase-locking.

The goal of this Letter is to show that type I SPOPOs in which the cavity is tuned to the first transverse mode family at the subharmonic are true noncritical sources of squeezed frequency combs above threshold, as well as perfect platforms for the study of spontaneous rotational symmetry breaking.

\textbf{Model.-- }As sketched in Fig. \ref{fig-Sketch}, we consider a type I SPOPO in which the cavity is transparent for the pump (an assumption that simplifies the analysis and is closer to current experiments). The pump comes from a mode-locked laser with Gaussian transverse profile, while the cavity is tuned so that only the first family of transverse modes is present at frequencies $\omega_{\mathrm{s},m}$, meaning that the down-converted photons can only take the profile of the lowest order Laguerre-Gauss modes with $\pm 1$ OAM. Assuming as usual that the resonator Rayleigh length is much larger than the crystal length, we can write the signal light electric field inside the nonlinear crystal as \cite{NavarretePhDthesis}
\begin{equation} \label{Esignal}
\hat{E}_\mathrm{s}=\mathrm{i}\mathcal{E}_\mathrm{s}\sum_{m\in\mathbb{Z}}\sum_{l=\pm 1}\hat{s}_{m,l}(t)L_{l}(\mathbf{r}_\bot)u_m(z) e^{-\mathrm{i}\omega_{\mathrm{s},m}t}+\mathrm{H.c.},
\end{equation}
where $\mathcal{E}_{\mathrm{s}}$ is the single photon field amplitude \cite{Patera10}, taken equal for all signal modes to an excellent approximation, $\hat{s}_{m,l}$ are the (interaction picture) annihilation operators for photons of frequency $\omega_{\mathrm{s},m}$ and transverse profile
\begin{equation} \label{uml}
L_{l}\left( \mathbf{r}_{\bot }\right) =\sqrt{\frac{2}{\pi }}\frac{r}{w_{\mathrm{s}}^{2}}e^{-r^{2}/w_{\mathrm{s}}^{2}}e^{\mathrm{i}l\phi },
\end{equation}
(OAM = $l$), with $\mathbf{r}_{\bot}=r(\cos\phi,\sin\phi)$ the transverse coordinates, $z$ is the axial coordinate, $w_\mathrm{s}$ is the spot size at the waist plane of the subharmonic modes ($z=0$ by definiteness), and $u_m(z)$ is the longitudinal shape of the mode, equal to $\exp(\mathrm{i}k_{\mathrm{s},m}z)$ for ring cavities or to $\sin[k_{\mathrm{s},m}(z+L_\mathrm{cav}/2)]$ for Fabry-Perot cavites \cite{Patera10}.

The quantum Heisenberg-Langevin equations describing the evolution of the operators $\hat{s}_{m,l}(t)$ are easily found by following the standard procedure explained in \cite{Patera10}, just taking into account that now PDC generates pairs of photons with opposite OAM, instead of spatially-degenerate pairs as in (\ref{Hint0}), which amounts for the replacement $\hat{s}_m^\dagger\hat{s}_q^\dagger\rightarrow\hat{s}_{m,+1}^\dagger\hat{s}_{q,-1}^\dagger$. We obtain
\begin{eqnarray}\label{LangevinEqs}
\frac{d\hat{s}_{m,l}}{dt}&=&-\gamma\hat{s}_{m,l}+\sqrt{2\gamma}\hat{s}_{\mathrm{in},m,l}(t)
\\
&&+\sum_{q}f_{m,q}[\gamma\sigma\alpha_{m+q}+\sqrt{2\kappa}\hat{p}_{\mathrm{in},m+q}(t)]\hat{s}_{q,-l}^\dag \notag
\\
&&-\kappa\sum_{j,q}f_{m,q}f_{j,m+q-j}\hat{s}_{j,+1}\hat{s}_{m+q-j,-1}\hat{s}_{q,-l}^\dag.\notag
\end{eqnarray}
Here, $\gamma$ is the decay rate through the partially transmitting mirror and $\kappa$ is the PDC two-photon damping rate, whose expressions in terms of physical parameters can be checked in \cite{Patera10}, and can be assumed equal for all modes. $\alpha_m$ are the normalized ($\sum_m\vert\alpha_m\vert^2=1$) spectral amplitudes of the pump frequency comb, and $\sigma=\sqrt{P/P_0}$, where $P$ is the external pump power and $P_0$ its value at the SPOPO threshold for monochromatic pumping ($\alpha_m=\delta_{m,0}$) \cite{Valcarcel06,Patera10}. Finally the ``in'' operators correspond to standard vacuum noise terms \cite{WallsMilburnBook,NavarretePhDthesis,GardinerBook}.

\textbf{Classical emission.--} The classical SPOPO dynamics is governed by Eqs. (\ref{LangevinEqs}) upon substituting operators $\hat{s}_{m,l}$ and $\hat{s}_{q,l}^\dag$ by complex variables $s_{m,l}$ and $s_{q,l }^\ast$, and ignoring vacuum noises. The solutions to the remaining nonlinear equations need to be evaluated numerically in general. They have however several general properties which will allow us to evaluate the most relevant quantum properties analytically.

First, there is the below-threshold solution, $s_{m,l}=0$ $\forall m$, which exists at any pumping level, but is unstable for $\sigma>\Lambda_0^{-1}$, where $\Lambda_0$ is the largest eigenvalue of the matrix $\mathcal{L}$ of elements $\mathcal{L}_{m,q}=f_{m,q}\alpha_{m+q}$, whose eigenvectors and eigenvalues $\Lambda_k$ define the squeezed below-threshold supermodes mentioned in the introduction. For $\sigma>\Lambda_0^{-1}$ a macroscopic field is built around the subharmonic frequency $\omega_0$, characterized by nonzero values of the classical spectral components, $s_{m,l}\neq 0$ in general. Since Eqs. (\ref{LangevinEqs}) have the symmetry $s_{m,\pm 1}\rightarrow e^{\mp \mathrm{i}\theta}s_{m,\pm 1}$ (with $\theta$ an arbitrary phase), the collective phase difference between opposite OAM modes is not fixed. On the other hand, if the pumping amplitudes $\alpha_m$ are real (non-chirped pulses), experimental and theoretical analysis on standard Gaussian SPOPOs \cite{Marandi14,Takata16,McMahon16,Inagaki16,Wang13,Haribara16,Hamerly16} have shown that there exist a large parameter region where the phases of the spectral components get locked to 0 or $\pi$. This carries on to the phase sums between opposite OAM modes in our model, leading to a stationary solution $\bar{s}_{m,\pm 1}=\rho_m e^{\mp\mathrm{i}\theta}$, with $\rho_m\in\mathbb{R}$ and $\theta$ an arbitrary phase. Eqs. (\ref{Esignal}) and (\ref{uml}) then provide a classical field
\begin{equation} \label{ClassE}
\bar{E}_\mathrm{s}(\mathbf{r},z,t)=\mathcal{E}_\mathrm{s} H_{10}(r,\phi-\theta)F(z,t),
\end{equation}
where $H_{10}(r,\phi-\theta)=\sqrt{\frac{8}{\pi}}w_\mathrm{s}^{-2} r e^{-r^2/w_\mathrm{s}^2}\cos(\phi-\theta)$ is a Hermite-Gauss, or TEM$_{10}$, transverse mode rotated by an angle $\theta$ with respect to the $x$ axis, and $F(z,t)=\sum_{m\in\mathbb{Z}}\rho_m \text{Im}\left\{u_m(z)e^{-\mathrm{i}\omega_{\mathrm{s},m}t}\right\}$. Hence, the spatiotemporal shape of the signal mean field emitted above threshold is the simple product of some propagating (wave) profile given by $F(z,t)$ (the phase-locked frequency comb) and a TEM$_{10}$ transverse spatial mode given by $H_{10}(r,\phi-\theta)$, which breaks the rotational symmetry of the system.

\textbf{Quantum properties of the emitted field.--} In previous works \cite{Navarrete08,Navarrete10,NavarretePhDthesis} we have studied how quantum noise affects the phase $\theta$ undefined at the classical level, proving that it diffuses linearly with time, driven by quantum noise. The analysis requires a rather technical procedure based the positive \textit{P} representation \cite{Drummond80}, and hence we present it in the supplemental material. Nevertheless, the presence of a noncritically squeezed mode in the system can be proven without resorting to such a rigorous analysis, and, for the sake of simplicity, in the following we just assume that above-threshold emission happens in a stable TEM$_{10}$ mode oriented within the $x$ axis ($\theta=0$), which we call the \textit{bright mode}. We will come back on this point in the last section.

In order to analyze the quantum properties of the down-converted frequency comb we then linearize the quantum Langevin equations (\ref{LangevinEqs}) around the classical solution $\bar{s}_{m,\pm 1}$ with $\theta=0$. It is quite remarkable that, as we show now, all the properties related to spontaneous symmetry breaking can be determined analytically without the need of specifying the steady-state amplitudes $\{\rho_m\}_{m\in\mathbb{Z}}$. In particular, we are interested in proving that the mode spatially orthogonal to the bright one, that is, the TEM$_{01}$ mode with the same temporal profile $F(z,t)$ as (\ref{ClassE}) but rotated by 90 degrees on the transverse plane (which we call \textit{dark mode}), has perfect quadrature squeezing irrespective of the distance to threshold.

Let us introduce the \textit{horizontal} and \textit{vertical} Hermite-Gauss annihilation operators $\hat{s}_{m,\mathrm{h}}=(\hat{s}_{m,+1}+\hat{s}_{m,-1})/\sqrt{2}$ and $\hat{s}_{m,\mathrm{v}}=\mathrm{i}(\hat{s}_{m,+1}-\hat{s}_{m,-1})/\sqrt{2}$, corresponding to TEM$_{10}$ (h) and TEM$_{01}$ (v) transverse modes respectively, and linearize the quantum Langevin equations (\ref{LangevinEqs}) with respect to the quantum fluctuations $\delta\hat{s}_{m,\mathrm{h}}=\hat{s}_{m,\mathrm{h}}-\rho_m$ and $\delta\hat{s}_{m,\mathrm{v}}=\hat{s}_{m,\mathrm{v}}$. The dynamics of the horizontal and vertical subspaces decouple in the linear approximation, and we get for the vertical subspace:
\begin{equation} \label{LinLanV}
\frac{d}{dt}\delta\hat{\mathbf{s}}_\mathrm{v}=\mathcal{L}_\mathrm{v}\delta\mathbf{\hat{s}}_\mathrm{v}+\sqrt{2\gamma}\mathbf{\hat{s}}_\mathrm{in,v}(t),
\end{equation}
where $\mathbf{\hat{s}}_{\mathrm{v}}=\text{col}(...,\hat{s}_{j,\mathrm{v}},...,\hat{s}_{j,\mathrm{v}}^\dag,...)$ and similarly for the vertical input noises $\hat{s}_{\mathrm{in},m,\mathrm{v}}=\mathrm{i}(\hat{s}_{\mathrm{in,}m,+1}-\hat{s}_{\mathrm{in,}m,-1})/\sqrt{2}$,
\begin{equation} \label{Lv}
\mathcal{L}_{\mathrm{v}}=\left(\begin{array}{cc}
-\gamma\mathcal{I} & \mathcal{R}
\\
\mathcal{R} & -\gamma\mathcal{I}%
\end{array}\right),
\end{equation}
with $\mathcal{I}$ the identity matrix with the proper dimensions, and the real, symmetric matrix $\mathcal{R}$ has elements
\begin{equation} \label{R}
\mathcal{R}_{m,q}=\gamma\sigma f_{m,q}\alpha_{m+q}-\kappa\sum_n f_{m,q}f_{n,m+q-n}\rho_n\rho_{m+q-n}.
\end{equation}
From the classical steady-state equation, it follows that $\boldsymbol{\rho}=\text{col}(...,\rho_{-1},\rho_0,\rho_{+1},...)$ is an eigenvector of $\mathcal{R}$ with $\gamma$ eigenvalue, and hence, $\mathbf{w}_1=\text{col}(\boldsymbol{\rho},-\boldsymbol{\rho})$ is an eigenvector of the full matrix $\mathcal{L}_\mathrm{v}$ with $-2\gamma$ eigenvalue. On the other hand, let us define $\hat{Y}_\mathrm{d}=\mathrm{i}\vert\boldsymbol{\rho}\vert^{-1}\sum_m\rho_m(\hat{s}_{m,\mathrm{v}}^\dag-\hat{s}_{m,\mathrm{v}})$, which corresponds to the quadrature measured in a homodyne detection with local oscillator matching the dark mode and $\pi/2$ shifted with respect to the pump beam. Then, note that $\mathbf{w}_1^T\delta\mathbf{\hat{s}}_\mathrm{v}(t)=\mathrm{i}\vert\boldsymbol{\rho}\vert\delta \hat{Y}_\mathrm{d}(t)$, and hence, projecting (\ref{LinLanV}) onto $\mathbf{w}_1$, we find
\begin{equation} \label{YdLinEq}
\frac{d}{dt}\delta\hat{Y}_\mathrm{d}=-2\gamma \delta\hat{Y}_\mathrm{d} - \mathrm{i}\sqrt{2\gamma/\vert\boldsymbol{\rho}\vert^2}\mathbf{w}_1^T\mathbf{\hat{s}}_\mathrm{in,v}(t).
\end{equation}
The relevant object in experiments and most applications is the noise spectrum \cite{WallsMilburnBook,NavarretePhDthesis}
\begin{equation} \label{VoutYd}
V_{Y_\mathrm{d}}^\mathrm{out}(\omega)=\int_{-\infty }^{+\infty }\hspace{-2mm}d\tau e^{-\mathrm{i}\omega\tau}\lim_{t\rightarrow +\infty}\langle\delta\hat{Y}_\mathrm{d,out}(t)\delta\hat{Y}_\mathrm{d,out}(t+\tau)\rangle,
\end{equation}
of the output quadrature $\delta\hat{Y}_\mathrm{d,out}=\sqrt{2\gamma}\delta\hat{Y}_\mathrm{d}-\delta\hat{Y}_\mathrm{d,in}$, with $\delta\hat{Y}_\mathrm{d,in}=-\mathrm{i}\mathbf{w}_1^T\mathbf{\hat{s}}_\mathrm{in,v}(t)/\vert\boldsymbol{\rho}\vert$, which measures the homodyne spectral noise power at frequency $\omega$, and signals squeezing whenever it is below 1 (0 meaning no noise: perfect squeezing). From the linear equation (\ref{YdLinEq}), it is straightforward to find
\begin{equation} \label{VoutYd2}
V_{Y_\mathrm{d}}^\mathrm{out}(\omega)=1-[1+(\omega/2\gamma)^2]^{-1},
\end{equation}
proving that the quadrature $\hat{Y}_\mathrm{d}$ has perfect squeezing at zero noise frequency \textit{irrespective of the system parameters}, that is, it shows perfect noncritical squeezing as we wanted to prove.

\textbf{Discussion}.-- We have shown that type I SPOPOs pumped by mode-locked laser of Gaussian transverse profile, but tuned at the subharmonic frequencies to the first transverse mode family, are the perfect platform where studying the consequences that spontaneous symmetry breaking has on the quantum state of nonlinear optical cavities \cite{Perez06,Perez07,Navarrete08,Navarrete09,Garcia09,Navarrete10,Garcia10,Navarrete11,Navarrete13,NavarretePhDthesis}. In particular, such device will emit a (classical) frequency comb with the transverse profile of a TEM$_{10}$ mode (bright mode), and a perfectly squeezed mode with the same spectral profile but an orthogonal TEM$_{01}$ spatial profile (dark mode). As shown in previous works \cite{Navarrete08,Navarrete10}, such perfect squeezing must be facilitated by the complete quantum indeterminacy of the bright and dark modes' orientation $\theta$. Unfortunately, within the familiar picture of quantum Langevin equations in which operators evolve, keeping track of the quantum dynamics of such a phase is highly non trivial (since the corresponding operator is quite intricate \cite{Luis93,Yu97}), and a consistent linearization procedure including the phase dynamics has to be performed within a stochastic representation. We develop explicitly such a technical approach in the supplemental material, where we obtain a variance of $\theta$ given by $\gamma t/4\vert \boldsymbol{\rho}\vert^2$. Hence, the modes' orientation diffuses with time, but at a slower rate the further we are from threshold (as $\vert\boldsymbol{\rho}\vert^2$ gets larger). In previous works we proved that a fixed local oscillator which does not follow the rotation of the dark mode is therefore still capable of measuring high levels of squeezing \cite{Navarrete10}, and, furthermore, either the injection of a weak seed with the spatiotemporal shape of the bright mode or a small anisotropy in the system are able to fix the orientation of the latter without degrading too much the squeezing of the dark mode \cite{Navarrete11,Navarrete13}. All these properties extend naturally to our proposed SPOPO configuration, which will then allow access to this elusive and rich regime of operation where spontaneous symmetry breaking dominates the quantum properties of the output light.

\begin{acknowledgments}
We thank Eugenio Rold\'{a}n and Nicolas Treps for fruitful discussions. G.P. and G.J.V. thank the Theory Division of the Max-Planck Institute of Quantum Optics, where the initial steps of the project were given, for their hospitality. This work has been supported by the Spanish Government and the European Union FEDER through projects FIS2011-26960 and FIS2014-60715-P. C.N.-B. acknowledges the financial support of the European Commission's project MALICIA (FP7-ICT-265522), and of the Alexander von Humboldt Foundation through its Fellowship for Postdoctoral Researchers.
\end{acknowledgments}


\appendix

\section{Supplemental material}

In this section we analyze the quantum properties of the SPOPO including the quantum diffusion of the orientation $\theta$, what we do rigorously by using the stochastic equations of the system within the positive \textit{P} representation \cite{Drummond80}. To this aim we first derive a master equation for our SPOPO configuration. Then, passing through a Fokker-Planck equation, we will retrieve the corresponding stochastic Langevin equations. Finally we will generalize the linearization technique developed in \cite{Navarrete08,Navarrete10} to discuss analytically the dynamics of the orientation $\theta$ and the quantum fluctuations of the dark mode.

\subsubsection{Master equation}

Our starting point is the Hamiltonian describing the interaction of the signal cavity resonances with the continuum of modes outside the cavity (through the partially transmitting mirror) and the continuum of modes around the pump comb (through the nonlinear crystal). Assuming that each spectral line interacts independently with its own reservoir (a good approximation as long as the cavity free spectral range $\Omega$ is larger than the couplings to the reservoir, a very good approximation in nonlinear cavity quantum optics), the Hamiltonian can be written as $\hat{H}=\hat{H}_\mathrm{s}+\hat{H}_\mathrm{p}+\hat{H}_\mathrm{b}+\hat{H}_\mathrm{ps}+\hat{H}_\mathrm{bs}$, with
\begin{subequations}
\begin{eqnarray}
\hat{H}_\mathrm{s}&=&\sum_{j,l}\hbar\omega_{\mathrm{s},j}\hat{s}^\dagger_{j,l}\hat{s}_{j,l},
\\
\hat{H}_\mathrm{b}&=&\sum_{j,l}\int_{\mathcal{O}(\omega_{\mathrm{s},j})}d\omega\hbar\omega\hat{b}^\dagger_{j,l}(\omega)\hat{b}_{j,l}(\omega),
\\
\hat{H}_\mathrm{p}&=&\sum_{j}\int_{\mathcal{O}(\omega_{\mathrm{p},j})}d\omega\hbar\omega\hat{p}^\dagger_j(\omega)\hat{p}_j(\omega),
\\
\hat{H}_\mathrm{bs}&=&\mathrm{i}\hbar\sqrt{\frac{\gamma}{\pi}}\sum_{j,l}\int_{\mathcal{O}(\omega_{\mathrm{s},j})}\hspace{-5mm}d\omega[\hat{b}_{j,l}(\omega)\hat{s}_{j,l}^\dagger-\hat{b}_{j,l}^\dagger(\omega)\hat{s}_{j,l}],
\\
\hat{H}_\mathrm{ps}&=&\mathrm{i}\hbar\sqrt{\frac{2\kappa}{\pi}}\sum_{jm}f_{j,m}\int_{\mathcal{O}(\omega_{\mathrm{p},j+m})}\hspace{-12mm}d\omega
\\
&&\hspace{0.5cm}\times[\hat{p}_{j+m}(\omega)\hat{s}_{j,+1}^\dagger\hat{s}_{m,-1}^\dagger-\hat{p}_{j+m}^\dagger(\omega)\hat{s}_{j,+1}\hat{s}_{m,-1}]\notag.
\end{eqnarray}
\end{subequations}
The first three terms account for the free evolution of the relevant modes; the second to last term describes the interconversion between external photons and cavity photons; the last term models the down-conversion process inside the crystal. All the commutators between the bosonic operators appearing in the expression are zero, except for $[\hat{s}_{j,l},\hat{s}^\dagger_{j',l'}]=\delta_{jj'}\delta_{ll'}$, $[\hat{b}_{j,l}(\omega),\hat{b}^\dagger_{j',l'}(\omega')]=\delta_{jj'}\delta_{ll'}\delta(\omega-\omega')$, and $[\hat{p}_j(\omega),\hat{p}^\dagger_{j'}(\omega')]=\delta_{jj'}\delta(\omega-\omega')$. All the parameters have been defined in the main text, while $\mathcal{O}(\omega)$ denotes a short spectral interval (smaller than $\Omega$) centered at $\omega$.

In the Heisenberg picture, a formal integration of the reservoir equations \cite{Patera10,Patera12} leads to the Heisenberg-Langevin equations (\ref{LangevinEqs}) used in the main text. In this section, however, we proceed in the Schr\"odinger picture where the state of the system evolves, and derive a master equation for the reduced state of the cavity modes. Before proceeding, it is convenient to move to a new picture defined by the transformation operators $\hat{U}_c=\exp(\hat{H}_c t/\mathrm{i}\hbar)$, with
\begin{align}
\hat{H}_c=&\sum_{j,l}\hbar\omega_{\mathrm{s},j}\hat{s}^\dagger_{j,l}\hat{s}_{j,l}
\\&+\sum_{j,l}\int_{\mathcal{O}(\omega_{\mathrm{s},j})}d\omega\hbar\omega_{\mathrm{s},j}\hat{b}^\dagger_{j,l}(\omega)\hat{b}_{j,l}(\omega)\notag
\\&+\sum_{j}\int_{\mathcal{O}(\omega_{\mathrm{p},j})}d\omega\hbar\omega_{\mathrm{p},j}\hat{p}^\dagger_j(\omega)\hat{p}_j(\omega)\notag
\end{align}
and
\begin{equation}
\hat{D}=\exp\left\{\sum_j\int_{\mathcal{O}(\omega_{\mathrm{p},j})}\hspace{-7mm}d\omega[\beta_j^*(\omega)\hat{p}_j(\omega)-\beta_j(\omega)\hat{p}^\dagger_j(\omega)]\right\}\hspace{-1mm},
\end{equation}
with $\beta_j(\omega)=\gamma\sigma\sqrt{\pi/2\kappa}\delta(\omega-\omega_{\mathrm{p},j})\alpha_j$. Note that the $\hat{D}$ displaces the pump field such that the coherent train of pulses (or frequency comb) injected in the cavity corresponds to vacuum in the new picture, where the state evolves according to the Hamiltonian $\hat{H}'=\hat{H}'_\mathrm{s}+\hat{H}'_\mathrm{p}+\hat{H}'_\mathrm{b}+\hat{H}_\mathrm{ps}+\hat{H}_\mathrm{bs}$, with
\begin{subequations}
\begin{eqnarray}
\hat{H}'_\mathrm{s}&=&\mathrm{i}\hbar\sqrt{\frac{2\kappa}{\pi}}\sum_{jm}f_{j,m}\int_{\mathcal{O}(\omega_{\mathrm{p},j+m})}\hspace{-12mm}d\omega
\\
&&\times[\beta_{j+m}(\omega)\hat{s}_{j,+1}^\dagger\hat{s}_{m,-1}^\dagger-\beta_{j+m}^\ast(\omega)\hat{s}_{j,+1}\hat{s}_{m,-1}]\notag,
\\
\hat{H}'_\mathrm{b}&=&\sum_{j,l}\int_{\mathcal{O}(\omega_{\mathrm{s},j})}d\omega\hbar(\omega-\omega_{\mathrm{s},j})\hat{b}^\dagger_{j,l}(\omega)\hat{b}_{j,l}(\omega),
\\
\hat{H}'_\mathrm{p}&=&\sum_{j}\int_{\mathcal{O}(\omega_{\mathrm{p},j})}d\omega\hbar(\omega-\omega_{\mathrm{p},j})\hat{p}^\dagger_j(\omega)\hat{p}_j(\omega).
\end{eqnarray}
\end{subequations}

In this picture, we can eliminate (trace out) the continuous reservoirs by taking vacuum as the reference state for all their modes and applying standard techniques \cite{GardinerBook}, arriving to the following master equation for the signal state $\hat{\rho}$:
\begin{align}  \label{me5}
\frac{d}{dt}\hat{\rho}&= \left[\gamma\sigma\sum_{j,m}f_{j,m}(\alpha_{j+m}^*\hat{s}_{j,+1}\hat{s}_{m,-1}-\mathrm{H.c.}),\hat{\rho}\right]
\\
&+\gamma\sum_{j,l}(2\hat{s}_{j,l}\hat{\rho}\hat{s}_{j,l}^\dag-\hat{s}_{j,l}^\dag\hat{s}_{j,l}\hat{\rho}-\hat{\rho}\hat{s}_{j,l}^\dag\hat{s}_{j,l})\notag
\\
&+\kappa\sum_{qmpn}f_{q,m}f_{p,n}\delta_{p+n,q+m}(2\hat{s}_{p,+1}\hat{s}_{n,-1}\hat{\rho}\hat{s}_{q,+1}^\dag \hat{s}_{m,-1}^\dag\notag
\\
&-\hat{s}_{p,+1}^\dag \hat{s}_{n,-1}^\dag \hat{s}_{q,+1}\hat{s}_{m,-1}\hat{\rho}-\hat{\rho}\hat{s}_{q,+1}^\dag \hat{s}_{m,-1}^\dag \hat{s}_{p,+1}s_{n,-1}). \notag
\end{align}

\subsubsection{Stochastic Langevin equations}

The master equation can be turned into an equivalent set of stochastic equations by following standard techniques based on the positive \textit{P} representation of the state \cite{Drummond80}. In our case, it is simple to show that such a distribution obeys the following Fokker-Planck equation:
\begin{equation}\label{FokkerPlanck}
\frac{\partial}{\partial t}P(\mathbf{s},\mathbf{s}^{+};t)= \hspace{-0.7mm}\left[-\sum_{i}\partial_i \mathcal{A}_i+ \frac{1}{2}\sum_{i,j}\partial_i\partial_j \mathcal{D}_{i,j} \right]\hspace{-1mm}P(\mathbf{s},\mathbf{s}^{+};t),
\end{equation}
where the indices $i$ and $j$ run over the set $\{\mathbf{s},\mathbf{s}^{+}\} $, with $\mathbf{s}=(...,s_{j,+1},...,s_{j,-1},...)$ and $\mathbf{s}^+=(...,s_{j,+1}^+,...,s_{j,-1}^+,...)$, and the components of the drift vector read
\begin{subequations}\label{driftmatrix}
\begin{eqnarray}
\mathcal{A}_{s_{m,l}}&=& -\gamma s_{m,l}+ \gamma\sigma\sum_q f_{m,q}\alpha_{m+q}
s_{q,-l}^+ +
\\
&&-\kappa\sum_{q,n,p}f_{m,q}f_{n,p}\delta_{m+q,n+p}s_{n,+1}s_{p,-1}s_{q,-l}^+,\notag
\\
\mathcal{A}_{s_{m,l}^+}&=& -\gamma s_{m,l}^+ + \gamma\sigma\sum_q f_{m,q}\alpha_{m+q}^*
s_{q,-l} +
\\
&&-\kappa\sum_{q,n,p}f_{m,q}f_{n,p}\delta_{m+q,n+p}s_{n,+1}^+s_{p,-1}^+s_{q,-l},\notag
\end{eqnarray}
\end{subequations}
while the elements of the diffusion matrix are found to be $\mathcal{D}_{s_{m,l},s_{q,k}^+}=0$, $\mathcal{D}_{s_{m,l}^+,s_{q,k}}=0$, $\mathcal{D}_{s_{m,l},s_{q,k}}=\delta_{l,-k}\mathcal{R}_{m,q;l}$, and $\mathcal{D}_{s_{m,l}^+,s_{q,k}^+}=\delta_{l,-k}\mathcal{R}_{m,q;l}^+$, with
\begin{subequations}\label{diffusionmatrix}
\begin{align}
\mathcal{R}_{m,q;l}&=\kappa\sum_{n,p}f_{m,q}f_{n,p}\delta_{m+q,n+p}s_{p,-1}s_{n,+1},
\\
\mathcal{R}_{m,q;l}^+&=\kappa\sum_{n,p}f_{m,q}f_{n,p}\delta_{m+q,n+p}s_{p,-1}^+ s_{n,+1}^+.
\end{align}
\end{subequations}

In order to write down the stochastic equations associated to Eq. \eqref{FokkerPlanck}, we first need to find the noise matrix $\mathcal{B}$ satisfying $\mathcal{B}\mathcal{B}^{\mathrm{T}}=\mathcal{D}$ \cite{GardinerBook2}. Since the diffusion matrix can be written in the block form
\begin{equation}
\mathcal{D}= \left(
\begin{array}{cc}
D & 0 \\
0 & D^+%
\end{array}
\right)= \left(
\begin{array}{cccc}
0 & \mathcal{R} & 0 & 0 \\
\mathcal{R} & 0 & 0 & 0 \\
0 & 0 & 0 & \mathcal{R}^+ \\
0 & 0 & \mathcal{R}^+ & 0%
\end{array}
\right),
\end{equation}
so does the noise matrix
\begin{equation}
\mathcal{B}= \left(
\begin{array}{cc}
B & 0 \\
0 & B^+%
\end{array}
\right),
\end{equation}
such that $BB^{\mathrm{T}}=D$ and $B^+B^{+T}=D^+$. Assuming, for sake of argument, that the index of longitudinal modes runs from $-N$ to $N$, the matrix $D$ has dimension $2(2N+1)\times2(2N+1)$. On the contrary the noise matrix $B$ does not need to be square like $D$, its only constrain being that it has to be a $2(2N+1)\times d_{\mathrm{B}}$ matrix, where we call $d_{\mathrm{B}}$ its \textit{internal dimension}. Then we write the diffusion matrix as
\begin{equation}
D=\sum_{m,q=-N}^{N}\sum_{l=\pm1}D^{[m,q;l]},
\end{equation}
where $D^{[m,q;l]}$ is the $2(2N+1)\times 2(2N+1)$ diffusion matrix associated to the down-conversion of the pair of modes $(s_{m,l},s_{q,-l})$ such that
\begin{equation}
D^{[m,q;l]}_{i,j}=\left\{
\begin{array}{ll}
\mathcal{R}_{m,q} & \text{if } i=(m,l) \text{ and } j=(q,-l)
\\
\mathcal{R}_{m,q} & \text{if } i=(m,-l) \text{ and } j=(q,l)
\\
0 & \text{otherwise}
\end{array}
\right. .
\end{equation}
Such a matrix has a simple related noise matrix which can be written as
\begin{equation}
B^{[m,q;l]}=\sqrt{\frac{\mathcal{R}_{mq}}{2}}\left(
\begin{array}{cc}
0 & 0
\\
\vdots & \vdots
\\
1 & \mathrm{i}
\\
\vdots & \vdots
\\
1 & -\mathrm{i}
\\
\vdots & \vdots
\\
0 & 0
\end{array}\right)
\begin{array}{c}
\\
\vspace{-4mm}
\\
\leftarrow(m,+l)
\\
\vspace{2mm}
\\
\leftarrow(q,-l)
\\
\\
\end{array},
\end{equation}
so that $B^{[m,q;l]}B^{[m,q;l]T}=D^{[m,q;l]}$. The full noise matrix of dimensions $2(2N+1)\times 4(2N+1)^2$ is then built as
\begin{equation}
B=\left(
\begin{array}{cccc}
\mathbb{B}^{[-N,-N]} & \mathbb{B}^{[-N,-N+1]} & ... & \mathbb{B}^{[N,N]}%
\end{array}
\right) ,
\end{equation}
with
\begin{equation}
\mathbb{B}^{[m,q]}= \left(
\begin{array}{cc}
B^{[m,q;l=-1]} & B^{[m,q;l=+1]}
\end{array}
\right)
\end{equation}
which has dimensions $2(2N+1)\times 4$. By construction the matrix $B$ satisfies
\begin{equation}
BB^{\mathrm{T}}=\sum_{m,q=-N}^{N}\sum_{l=\pm1} B^{[m,q;l]}\left.B^{[m,q;l]}\right.^{\mathrm{T}}=D.
\end{equation}
Analogously we get same results for $B^{+}$ but with the exchange $s\leftrightarrow s^{+}$.

We are now in conditions of writing the stochastic Langevin equation corresponding to eq. \eqref{FokkerPlanck}
\begin{subequations}
\begin{align}
\frac{d\mathbf{s}}{dt} & =\mathbf{A}(\mathbf{s},\mathbf{s}^{+})+B\boldsymbol{%
\eta}(t),  \label{stoca1} \\
\frac{d\mathbf{s}^{+}}{dt} & =\mathbf{A}^{+}(\mathbf{s},\mathbf{s}^{+})+B^{+}%
\boldsymbol{\eta}^{+}(t),  \label{stoca2}
\end{align}
\end{subequations}
where we have defined the vectors $\mathbf{A}$ and $\mathbf{A}^+$ with corresponding elements $A_{m,l}=\mathcal{A}_{s_{m,l}}$ and $A_{m,l}^{+}=\mathcal{A}_{s_{m,l}^{+}}$, while the components of $\boldsymbol{\eta}$ and $\boldsymbol{\eta}^{+}$ are independent real Gaussian white noises \cite{GardinerBook2}. Note that with our choice of noise matrix, we have to deal with $8(2N+1)^{2}$ noises, way above the minimal choice $4(2N+1)$, which might be bad for numerical purposes, but will make no difference for our linearized analytic approach.

\subsubsection{Linearization in the presence of \\spontaneous symmetry breaking}

In the following, we apply the linearization technique to the previous Langevin equations, which is an approximate method that we proved to lead to the correct predictions when working sufficiently above threshold \cite{Navarrete10}. The standard method proceeds by writing the stochastic amplitudes as $\mathbf{s}=\mathbf{\bar{s}}+\delta\mathbf{s}$ and $\mathbf{s}^{+}=\mathbf{\bar{s}}^{\ast}+\delta\mathbf{s}^{+}$, where $\mathbf{\bar{s}}$ is the classical stationary solution, that is, $\mathbf{A}(\mathbf{\bar{s}},\mathbf{\bar{s}}^{\ast})=0$; then, one assumes that the fluctuations $(\delta\mathbf{s},\delta\mathbf{s}^{+})$ and the noises $(\boldsymbol{\eta},\boldsymbol{\eta}^{+})$ are small, and therefore only terms up to linear on these must be considered. However, in our case the equations are invariant under changes of the phase difference between the Laguerre-Gauss modes, $\theta$, what means that there is a direction in phase space in which fluctuations are not damped, and hence they cannot be assumed small. This is the main reason why we use this positive \textit{P} formalism, since the linearization can still be performed by taking into account the fluctuations of the phase $\theta$ explicitly (which is not clear how to do in the Heisenberg picture, where the phase-difference operator has a very complicated expression \cite{Luis93,Yu97}). Let us now then introduce the proper linearization procedure for this case in which a continuous symmetry is broken \cite{Navarrete08,Navarrete10,Lane88,Reid88,Reid89}. In our case where the pump amplitudes $\alpha_j$ are real, and the classical solution has the form introduced in the main text, we proceed by writing the stochastic variables as
\begin{subequations}\label{StoBF}
\begin{align}
s_{m,\pm1}&=\left[\rho_{m}+b_{m,\pm1}(t)\right]e^{\mp\mathrm{i}\theta(t)},
\\
s_{m,\pm1}^{+}&=\left[\rho_{m}+b_{m,\pm1}^{+}(t)\right]e^{\pm\mathrm{i}%
\theta (t)},
\end{align}
\end{subequations}
where the phase $\theta(t)$ is taken as an explicit stochastic variable whose fluctuations account for the quantum fluctuations of the corresponding phase difference operator. Note that the classical amplitudes $\boldsymbol{\rho}$ satisfy the equation
\begin{equation}  \label{RhoVecEqs}
R\boldsymbol{\rho}=\gamma\boldsymbol{\rho},
\end{equation}
where we have defined the matrix
\begin{equation}
R_{m,q}=\gamma\sigma f_{m,q}\alpha_{m+q} -\kappa\sum_{n,p} f_{m,q}f_{n,p}\delta_{m+q,n+p}\rho_n\rho_p.
\end{equation}
This property will be of use later.

Writing the stochastic amplitudes in this way, we can now assume that the fluctuations
\begin{subequations}
\begin{eqnarray}
\mathbf{b}&=&\text{col}(...,b_{j,+1},...,b_{j,-1},...),
\\
\mathbf{b}^+&=&\text{col}(...,b_{j,+1}^+,...,b_{j,-1}^+,...),
\end{eqnarray}	
\end{subequations}
as well as the derivative of the phase $\dot{\theta}$, are of the order of the noises, while the phase $\theta$ itself is not bounded. This allows us to linearize the stochastic Langevin equations as
\begin{equation}\label{LinLanGen}
-\mathrm{i}\mathbf{u}_{0}\dot{\theta}+\mathbf{\dot{c}}=(\mathcal{L}-\gamma\mathcal{I})\mathbf{c}+\mathcal{F\bar{B}}\boldsymbol{\xi},
\end{equation}
where we have defined the vector of fluctuations $\mathbf{c}=\mathrm{col}(\mathbf{b},\mathbf{b}^+)$, the noise vector $\boldsymbol{\xi}=\mathrm{col}(\boldsymbol{\eta},\boldsymbol{\eta}^+)$, the vector $\mathbf{u}_{0}=\text{col}(\boldsymbol{\rho},-\boldsymbol{\rho},-\boldsymbol{\rho},\boldsymbol{\rho})$, and the matrices
\begin{equation}
\mathcal{L}=\left(
\begin{array}{cccc}
T & T & 0 & R \\
T & T & R & 0 \\
0 & R & T & T \\
R & 0 & T & T%
\end{array}
\right),
\end{equation}
\begin{equation}
\mathcal{\bar{B}}=\mathcal{B}(\mathbf{s}=\mathbf{\bar {s}},\mathbf{s}^{+}=%
\mathbf{\bar{s}}^{\ast}),
\end{equation}
and
\begin{equation}
\mathcal{F}=\left(
\begin{array}{cccc}
F & 0 & 0 & 0 \\
0 & F^{\ast} & 0 & 0 \\
0 & 0 & F^{\ast} & 0 \\
0 & 0 & 0 & F%
\end{array}
\right) ,
\end{equation}
with
\begin{equation}
T_{m,n}=-\kappa\sum_{q,p}f_{m,q}f_{n,p}\delta_{m+q,n+p}\rho_n\rho_p,
\end{equation}
and $F=e^{\mathrm{i}\theta}\mathcal{I}_{(2N+1)\times(2N+1)}$ is proportional to the identity of the proper dimension. Now, note that the Fokker-Planck equation associated to this stochastic equations is independent of $\theta$ \cite{NavarretePhDthesis}, and hence, we can take $\mathcal{F}=\mathcal{I}$ without loss of generality.

\subsubsection{Phase diffusion}

Using (\ref{RhoVecEqs}) and the various definitions above, it is easy to show that $\mathbf{u}_{0}$ is an eigenvector of $\mathcal{L}-\gamma\mathcal{I}$ with $0$ eigenvalue, that is, it is the \textit{Goldstone mode} linked to the symmetry of the system. Projecting the linearized equations onto $\mathbf{u}_{0}$, we get
\begin{equation}
\dot{\theta}=\frac{\mathrm{i}}{4|\boldsymbol{\rho}|^2}\mathbf{u}_{0}^{T}\mathcal{\bar{B}}\boldsymbol{\xi}(t),
\end{equation}
where we have used $\mathbf{u}_{0}^{T}\mathbf{u}_{0}=4|\boldsymbol{\rho}|^2$, and we have set $\mathbf{u}_{0}^{T}\mathbf{c}=0$ to remove the variable-redundancy that we introduced when writing the stochastic amplitudes as (\ref{StoBF}). This equation tells us that, as expected, the phase $\theta$ is solely driven by quantum noise. Its solution is
\begin{equation}
\theta(t)=\theta(0)+\frac{\mathrm{i}}{4|\boldsymbol{\rho}|^2}\int_{0}^{t}dt^{\prime }%
\mathbf{u}_{0}^{T}\mathcal{\bar{B}}\boldsymbol{\xi}(t^{\prime}),
\end{equation}
leading to a phase variance
\begin{equation}
V_{\theta}(t)=\langle[\theta(t)-\theta(0)]^2\rangle =-\frac{\mathbf{u}_{0}^{T}\mathcal{\bar{B}\bar{B}}^{T}\mathbf{u}_{0}}{%
16|\boldsymbol{\rho}|^4}t.
\end{equation}
Finally, using the fact that $\mathcal{\bar{B}\bar{B}}^{T}=\mathcal{\bar{D}}$, with
\begin{equation}
\mathcal{\bar{D}=\mathcal{D}}(\mathbf{s}=\mathbf{\bar{s}},\mathbf{s}^{+}=%
\mathbf{\bar{s}}^{\ast})\mathcal{=}\left(
\begin{array}{cccc}
0 & R & 0 & 0 \\
R & 0 & 0 & 0 \\
0 & 0 & 0 & R \\
0 & 0 & R & 0%
\end{array}
\right) ,
\end{equation}
so that $\mathbf{u}_{0}^{T}\mathcal{\bar{D}}\mathbf{u}_{0}=-4\boldsymbol{\rho }^{T}R\boldsymbol{\rho}=-4\gamma|\boldsymbol{\rho}|^2$, where we used (\ref{RhoVecEqs}), we get a phase variance which increases linearly with time as $V_{\theta}(t)=\gamma t/4|\boldsymbol{\rho}|^2$, just as was introduced in the main text.

\subsubsection{Quadrature fluctuations of the dark mode}

Let us consider the mode with the same temporal profile as the bright mode generated classically, but in a TEM$_{01}$ mode spatially orthogonal to the bright TEM$_{10}$ one. We referred to this as the \textit{dark mode}, and its corresponding stochastic amplitudes within the positive \textit{P} representation are given by
\begin{subequations}\label{DarkAmplitudes}
\begin{eqnarray}
s_\mathrm{d}&=&\frac{\mathrm{i}}{\sqrt{2}|\boldsymbol{\rho}|}\sum_n\rho_n\left(e^{\mathrm{i}\theta}s_{n,+1}-e^{-\mathrm{i}\theta}s_{n,-1}\right),
\\
s_\mathrm{d}^+&=&\frac{-\mathrm{i}}{\sqrt{2}|\boldsymbol{\rho}|}\sum_n\rho_n\left(e^{\mathrm{i}\theta}s_{n,+1}^+-e^{-\mathrm{i}\theta}s_{n,-1}^+\right).
\end{eqnarray}
\end{subequations}
In the following, we will prove that an output quadrature of this mode is perfectly squeezed at any pump level above threshold. To this aim, let us evaluate next the noise spectrum associated the quadratures $X_\mathrm{d}=s_\mathrm{d}^+ + s_\mathrm{d}$ and $Y_\mathrm{d}=\mathrm{i}(s_\mathrm{d}^+ - s_\mathrm{d})$, which can be written in terms of stochastic correlators as \cite{NavarretePhDthesis}
\begin{equation}\label{VoutStoch}
V^\mathrm{out}_{Q_\mathrm{d}}(\omega)=1+2\gamma\int_{-\infty}^{+\infty}d\tau e^{-\mathrm{i}\omega t} \lim_{t\rightarrow\infty}\langle Q_\mathrm{d}(t)Q_\mathrm{d}(t+\tau) \rangle,
\end{equation}
with $Q_\mathrm{d}=X_\mathrm{d},Y_\mathrm{d}$.

It is simple to show from (\ref{RhoVecEqs}) again that the vector $\mathbf{u}_1=\text{col}(\boldsymbol{\rho},-\boldsymbol{\rho},\boldsymbol{\rho},-\boldsymbol{\rho})$ is another eigenvector of the linear stability matrix $\mathcal{L}-\gamma\mathcal{I}$ with $-2\gamma$ eigenvalue. On the other hand, using (\ref{StoBF}) and (\ref{DarkAmplitudes}), we find the relations $\mathbf{u}_0^T\mathbf{c}=-\mathrm{i}\sqrt{2}|\boldsymbol{\rho}|X_\mathrm{d}$ and $\mathbf{u}_1^T\mathbf{c}=\sqrt{2}|\boldsymbol{\rho}|Y_\mathrm{d}$. From the previous section, we then see that $X_\mathrm{d}(t)=0$, while projecting (\ref{LinLanGen}) onto $\mathbf{u}_1$, we obtain the following evolution equation for $Y_\mathrm{d}(t)$:
\begin{equation}
\dot{Y}_\mathrm{d}=-2\gamma Y_\mathrm{d}+\frac{1}{\sqrt{2}|\boldsymbol{\rho}|}\mathbf{u}_1^T\bar{\mathcal{B}}\boldsymbol{\xi}(t),
\end{equation}
leading to the two-time correlator
\begin{equation}
\lim_{t\rightarrow\infty}\langle Y_\mathrm{d}(t_1)Y_\mathrm{d}(t_2) \rangle=\frac{\mathbf{u}_1^T\bar{\mathcal{D}}\mathbf{u}_1}{8\gamma|\boldsymbol{\rho}|^2}e^{-2\gamma|t_1-t_2|}.
\end{equation}
Using next the property $\mathbf{u}_{0}^{T}\mathcal{\bar{D}}\mathbf{u}_{0}=-4\gamma|\boldsymbol{\rho}|^2$, and performing the Fourier transform appearing in (\ref{VoutStoch}), we finally obtain
\begin{subequations}
\begin{eqnarray}
V^\mathrm{out}_{Y_\mathrm{d}}(\omega)&=&1-[1+(\omega/2\gamma)^2]^{-1},
\\
V^\mathrm{out}_{X_\mathrm{d}}(\omega)&=&1,
\end{eqnarray}	
\end{subequations}
showing that, irrespectively of the system parameters, $Y_\mathrm{d}$ is perfectly squeezed at zero noise frequency, while $Q_\mathrm{d}$ has vacuum fluctuations at all noise frequencies.

\end{document}